\documentclass[aps,prb,superscriptaddress,prb,floatfix,twocolumn,amsmath,amssymb,showpacs]{revtex4}
\usepackage{graphicx}
\usepackage{dcolumn}
\usepackage{bm}
\begin{document}
\title{\bf
Thermal rippling behavior of graphane
}

\author{S. Costamagna}
\affiliation{Universiteit Antwerpen, Department of Physics, Groenenborgerlaan 171, BE-2020 Antwerpen, Belgium.
}
\affiliation{
Facultad de Ciencias Exactas Ingenier{\'\i}a y Agrimensura, Universidad Nacional de Rosario and Instituto de
F\'{\i}sica Rosario, Bv. 27 de Febrero 210 bis, 2000 Rosario,
Argentina.}

\author{M. Neek-Amal\footnote{corresponding author: neekamal@srttu.edu}}

\affiliation{Universiteit Antwerpen, Department of Physics, Groenenborgerlaan 171, BE-2020 Antwerpen, Belgium.
}

\author{J. H. Los}
\affiliation{Institute of Physical Chemistry and
Center for Computational Sciences, Johannes Gutenberg University Mainz,
Staudinger Weg 9, D-55128 Mainz, Germany}

\author{F. M. Peeters}
\affiliation{Universiteit Antwerpen, Department of Physics, Groenenborgerlaan 171, BE-2020 Antwerpen, Belgium.
}

\date{\today}

\begin{abstract}
Thermal fluctuations of single layer hydrogenated graphene
(graphane) are investigated using large scale atomistic simulations.
By analyzing the mean square value of the height fluctuations $\langle
h^2\rangle$ and the height-height correlation function $H(q)$ for
different system sizes and temperatures we show that hydrogenated
graphene is an un-rippled system in contrast to graphene.
The height fluctuations are bounded,
which is confirmed by a $ H(q) $ tending
to a constant in the long wavelength limit instead of showing the characteristic
scaling law $ q^{4-\eta}~(\eta \simeq 0.85)$ predicted by membrane theory.
This unexpected behaviour persists up to temperatures of at least 900 K
and is a consequence of the fact that in graphane
the thermal energy can be accommodated by in-plane bending modes, i.e.
modes involving C-C-C bond angles in the buckled carbon layer, instead
of leading to significant out-of-plane fluctuations
that occur in graphene.
\end{abstract}



\maketitle

\begin{figure}
\includegraphics[width=0.38\textwidth]{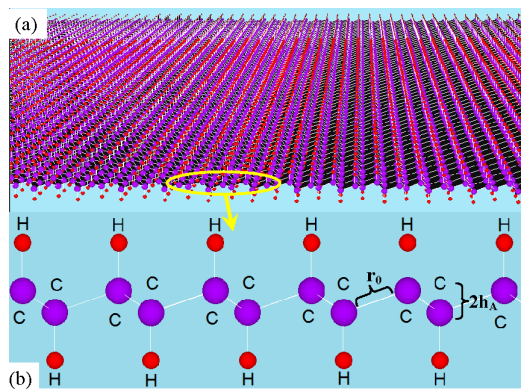}
\caption{(Color online) (a) Schematic view of a large sample of graphane.
(b) Buckling between C atoms in the A (higher) and B (lower) sublattices
at $T=300K$.
 \label{figGA} }
\end{figure}

Hydrogenated graphene
(GE), called graphane (GA), is a quasi two-dimensional (2D) structure
of carbon ($C$) atoms ordered in a buckled honey-comb lattice
covalently bonded to hydrogen ($H$) atoms in an alternating, chair-like
arrangement~\cite{sluiter-sofo}. Experimentally, it has been shown that
GA can be obtained reversibly starting from a pure GE layer~\cite{elias}
and since then it has become a material of high interest due to its
potential applications in nanoelectronics~\cite{int2}.
As compared to GE, the chemisorption of the $H$ atoms is accompanied by
an important reconstruction of the chemical bonds and angles in the
flat honeycomb lattice~\cite{mehdi}. Each carbon atom acquires an
$H$ neighbor, involving a transition from {\it sp$^2$} to {\it sp$^3$}
hybridization, which turn the conjugated, graphitic C-C bonds into single
C-C bonds changing locally the planar shape of graphene into
an angstrom scale out-of-plane buckled shaped membrane~\cite{PRB77}
as displayed schematically in Fig.~\ref{figGA}.

One expects that at non-zero temperature thermally excited ripples will distort the lattice.
For GE these ripples can be described theoretically by the
elasticity theory of continuum membranes~\cite{book2d}.
Amongst others, this membrane theory predicts a suppression of the
long wavelength out-of-plane fluctuations by the anharmonic coupling between the
out-of-plane bending and in-plane stretching modes leading to a characteristic
power law behavior for the height fluctuations as a function of the system size.
Although the height fluctuations still diverge, the normal-to-normal
correlation is preserved over a large lenght scale stabilizing the membrane.
By using atomistic simulations these results were shown to be applicable to
GE, implying that this prototype 2D solid behaves as a membrane~\cite{fas1,GE2d},
and also to bi-layer GE~\cite{bilayer}.
Experiments have confirmed that suspended GE is not perfectly flat but instead
presents ripples at finite temperatures~\cite{3}.

In this paper we study thermally excited ripples in GA
using state-of-the-art molecular dynamics (MD) simulations
and show that the angstrom scale thermal ripples present
in GE do not appear in GA for temperatures up to at
least 900 $K$.
%
The A- and B-sublattice buckling is
preserved and inhibits the formation of long wavelength ripples.
As a consequence, the calculated height-height correlation function $H(q)$
does not follow the characteristic $q^{4-\eta}$ power law scaling in the
long wavelength limit predicted by membrane theory, and the height
fluctuations appear to be bounded.

\vspace{0.1cm}
%
According to membrane theory and within the harmonic approximation,
applicable in the short wavelength regime where $ q $ is larger
than some cross-over value $ q^* $, the out-of-plane and in-plane modes
are decoupled and the elastic bending free energy density is
described by $ F_{harm} = \kappa (\nabla^2 h)^2 $, where $h$ is the
local height and $ \kappa $ is the bending rigidity of the membrane
which governs the properties of the temperature induced ripples.
Substitution of the
Fourier transform of $h$ and integrating over 2D space
leads to the following height-height correlation function
\begin{eqnarray}
H_{harm}(q)=\langle|h(q)|^2\rangle=\frac{N k_B T}{\kappa S_0 q^{4}}~~,
\label{hq1}
\end{eqnarray}
where $N$ is the number of atoms, $S_0$ the surface
area per atom and $k_B$ is the Boltzmann constant.
Accordingly, the height fluctuations in the harmonic regime behave as
$ \langle h^2 \rangle_{harm} = C L^2 $, with $C $ a temperature dependent
constant and $L$ the linear size of the system.
In the large wavelength limit, i.e. for $ q < q^* $, the height fluctuations
are suppressed by the mentioned anharmonic coupling between bending and
stretching modes giving rise to a renormalized $q$-dependent bending
rigidity $ \kappa_{R} \propto q^{-\eta} $ and a power law scaling behavior
\begin{eqnarray}
H(q)=\frac{N k_B T}{\kappa S_0 q^{4-\eta}}
\label{hq2}
\end{eqnarray}
and accordingly $ \langle h^2 \rangle = C'  L^{2-\eta} $
with $C'$ a constant ($\neq C)$. The universal scaling exponent
$\eta$ has been estimated to be $0.821$ \cite{ledousal}. For GE, using
MC simulations with the empirical LCBOPII potential \cite{LCBOPII},
good agreement with these results, derived from continuum theory,
was found with $\eta \approx 0.85$ [\onlinecite{fas1}].
Here, we investigate to which extent membrane theory can
be applied to the description of thermally excited ripples in GA and
we compare our results with those for GE.

\begin{figure}[t]
\includegraphics[width=0.45\textwidth]{fig2a.eps}
\vspace{0.17cm}

\includegraphics[width=0.4\textwidth]{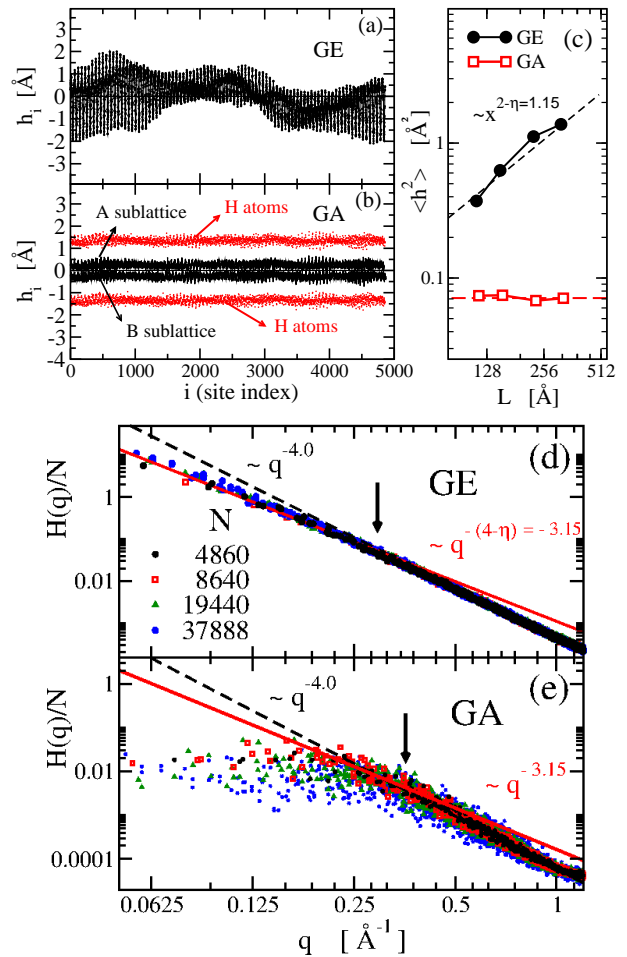}
\caption{(Color online) Heights of the C atoms in the GE (a) and GA (b)
against the site index for arbitrary snapshots taken during the MD simulation.
$N$=4860. $T$=300 $K$.
(c) $\langle h^2 \rangle$
against $L=\sqrt{L_x L_y}$ in GE (circles) and GA
(squares).
$H(q)$ for different system sizes as indicated for (d) GE and (e)
GA. The dashed line shows the harmonic $q^{-4}$ behavior
and the solid line the correction due to anharmonic coupling for small $q$.
Vertical arrows roughly indicate $q^*$ below which the harmonic behavior
is broken.
} \label{fig11}
\end{figure}

To calculate the height fluctuations for GA we first need to define
an appropriate value for the height $ h_i $ of each lattice site $i$.
Since we are mainly interested in the long wavelength fluctuations,
which normally govern the size of the height fluctuations,
we defined it on the basis of the carbon positions as
\begin{eqnarray*}
h_i = \frac{1}{2} \left( z_i + \frac{1}{3} \sum _{j}\nolimits'
z_j \right)
\label{hi}
\end{eqnarray*}
where $i$ is a carbon atom, $ \sum'_j $ runs over the three
carbon neighbors of $ i $ and $ z_i$ is the z-coordinate
perpendicular to the plane. This definition allows for a straightforward
comparison with GE, for which the heights were defined in the same way
following previous work \cite{bilayer}.
To measure $\langle h^2 \rangle$ and $H(q)$ we have performed MD
simulations using the LAMMPS package\cite{lammps}. Both the GA and
the GE systems have been sampled using the constant $NPT$ ensemble (with P=0).
For the interatomic interactions we used the modified second generation
of Brenner's bond-order potential, i.e. AIREBO\cite{AIREBO}, which has been shown
to predict correctly the configurations for many different hydrocarbon
structures.
We simulated square shaped systems with the
number of C atoms equal to $N=$ 4860, 8640, 19440 and 37888, in a temperature
range from $100-900~ K$ (note that for GA the total number of atoms is twice
large). Periodic boundary conditions were applied in the x- and y-directions.
The presented results have been computed averaging over $300-400$
uncorrelated configurations.
%

\begin{figure}
\includegraphics[width=0.4\textwidth]{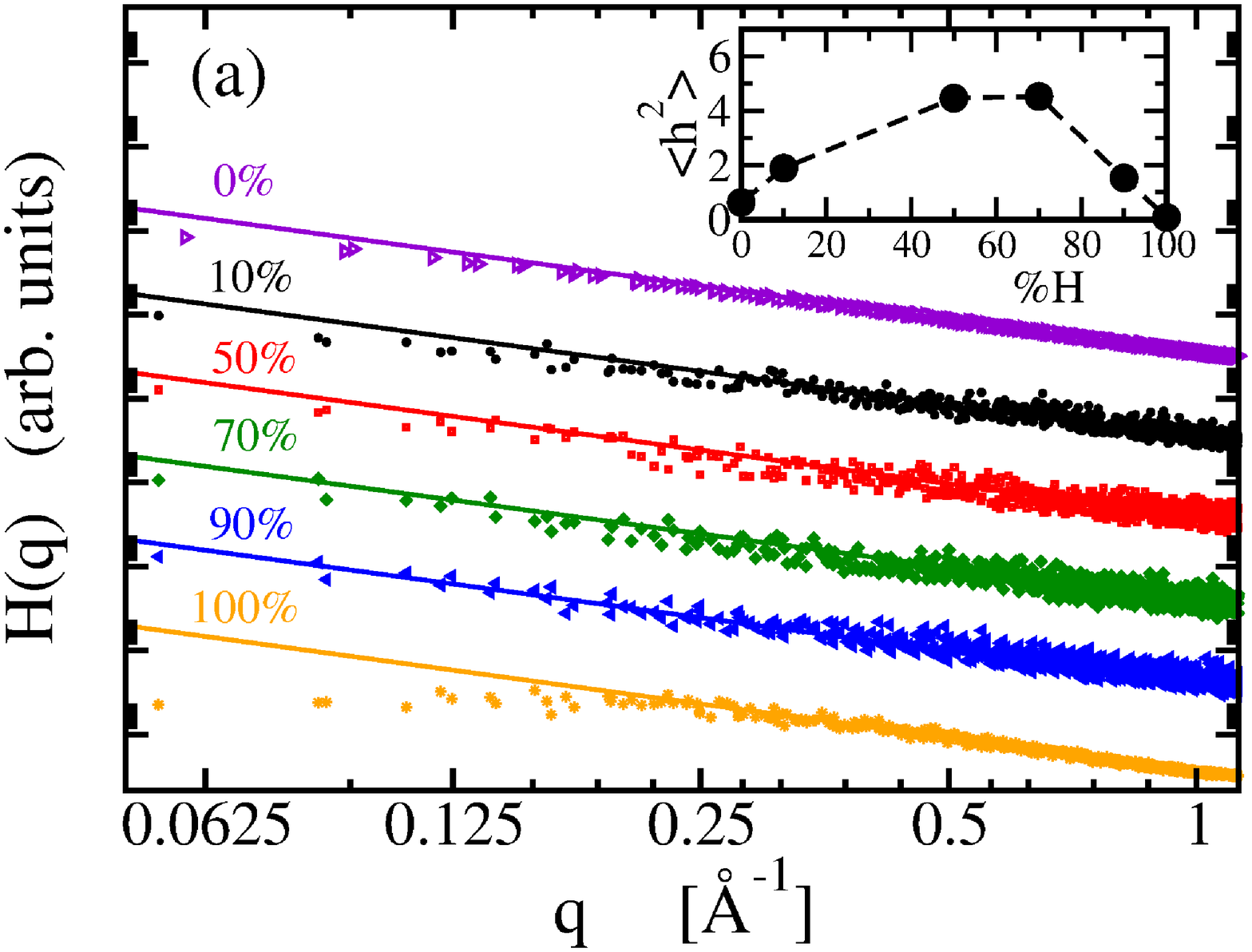}
\vspace{0.25cm}

\includegraphics[width=0.4\textwidth]{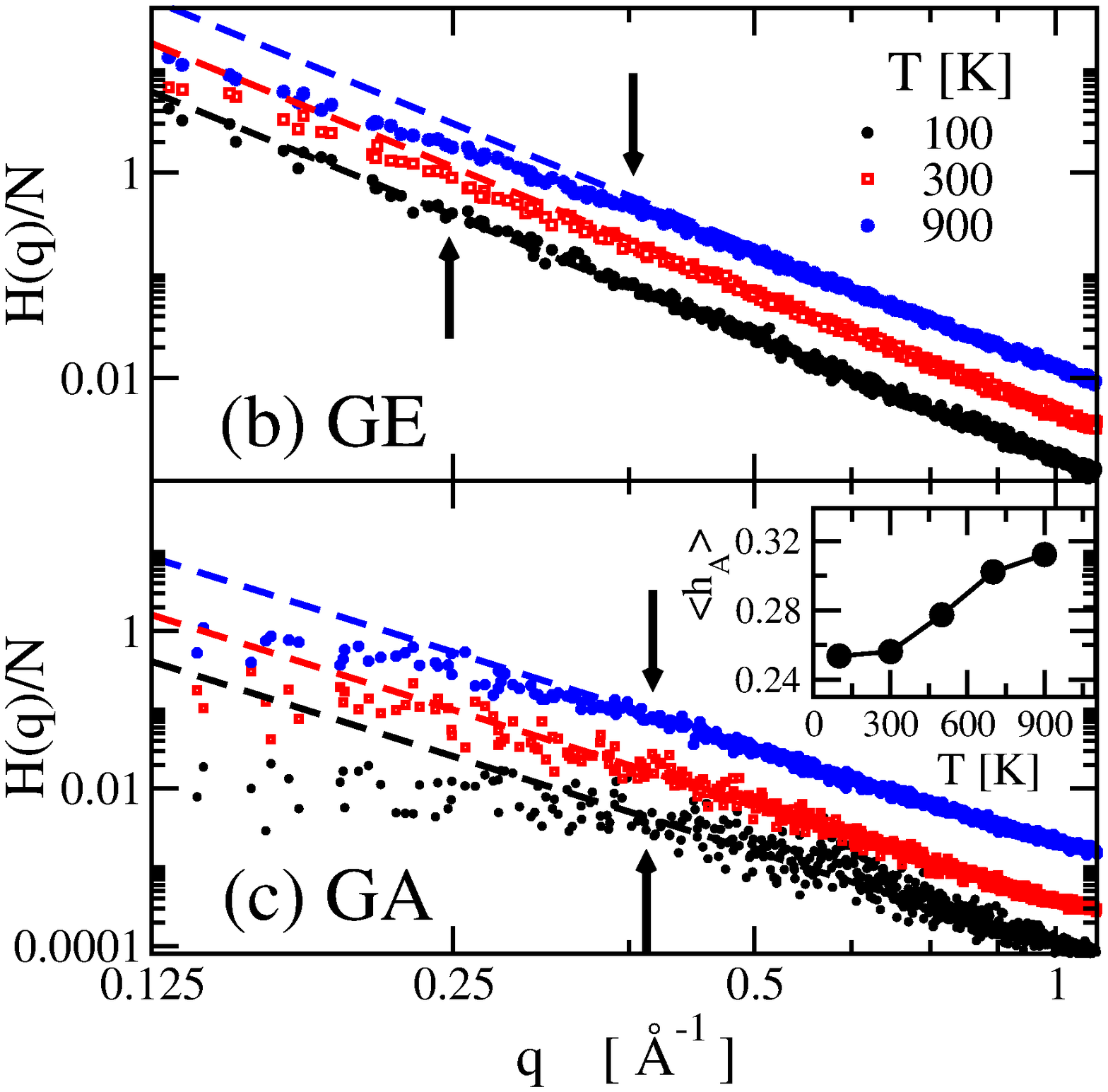}
\caption{(Color online)
(a) $H(q)$ for different $\%$ of H atoms
(0 $\%$ correspond to GE and 100 $\%$ to GA).
The different curves have been shifted for a better comparison.
In the inset we show the variation of $\langle h^2 \rangle$. $T=300K$.
$H(q)$ for different temperatures as indicated for (b) GE and (c) GA.
The inset of (c) shows the average value of the heights in
the A-sublattice of C atoms against temperature.
$N=8640$. } \vspace{0.15cm} \label{fig13}
\end{figure}

In Fig.~\ref{fig11} we show the out-of-plane positions for
GE (a) and GA (b)
for arbitrary snapshots taken
during the simulation at room temperature.
In GA, the A- and B-sublattices fluctuate around their mean heights
$ \langle h_{A,B}\rangle \cong \pm 0.256 \AA$.
The scaling of $\langle h^2 \rangle$  with the system size is
displayed in Fig.~\ref{fig11}(c).
The obtained values for GE are in close agreement with previous
reported MD results obtained with the REBO potential~\cite{liu} and
slightly lower than those obtained from MC simulations using the LCBOPII~\cite{fas1}.
For GE, $\langle h^2 \rangle$ increases as
$ L^{2-\eta} $ as expected from membrane theory.
For GA, instead, $\langle h^2 \rangle$ is almost independent
of the system size.
The differences in the rippling behavior of GA and GE are also
evident from the results for $H(q)$ (Figs.~\ref{fig11}(d, e)).
As it should be, the $ H(q) $ functions for different system sizes
overlap. However, for GA, although the harmonic $ q^{-4} $ behavior
for short wavelengths is well recovered, $ H(q) $ tends to a constant
in the long wavelength limit. Hence, it does not follow the $q^{4-\eta}$
power law as expected from membrane theory and found for GE.


The intermediate regimes where GE is only partially covered by H atoms are analyzed
in Fig.~\ref{fig13}(a).
Notice that $H(q)$ displays harmonic behavior even for a H covering as large as 90$\%$.
The deviations at low wave vectors are small
and hence can be ascribed by anharmonic coupling.
The variation of $\langle h^2\rangle$, shown in the inset,
indicates that first the sheet is softened when partially hydrogenated
and becomes stiff at full coverage.


The behavior of $H(q)$ for different temperatures is shown in
Fig.~\ref{fig13}(b) and (c) .
As expected, with increasing temperature $H(q)$ is
shifted to larger values for both GE and GA.
In the inset of Fig.~\ref{fig13}(c) we show the average heights
of the C atoms in the A-sublattice against temperature. We also found
that $\langle h_A\rangle \cong -\langle h_B\rangle$ over the whole
temperature range, implying that the A- and B-sublattice buckling
is preserved.

\begin{figure}[t]
\includegraphics[width=0.4\textwidth]{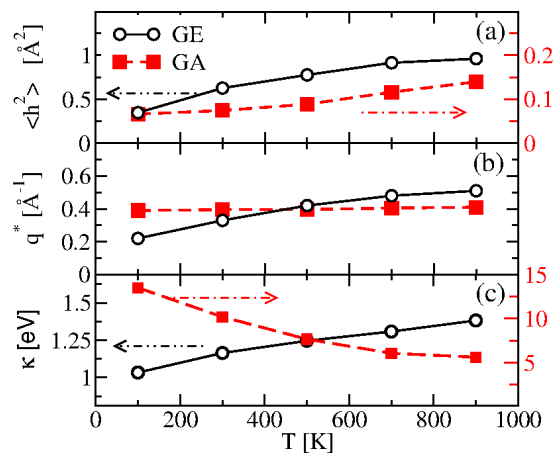}
\caption{Variation of (a) $\langle h^2 \rangle$,
(b) $q^*$ and (c) $\kappa$ against temperature for GE (circles) and GA (squares).
$N=8640$.} \label{fig14}
\end{figure}

More signals of the differences between the corrugations in GA and GE
comes from the temperature dependence of $\langle h^2 \rangle$ shown
in Fig. \ref{fig14}(a). Note the difference in vertical scale
displayed on the left- and right-hand side for GE and GA, respectively.
While in GE the value of $\langle h^2 \rangle$ changes about 0.61
\AA~ between 100 and 900 $K$ the variation is only 0.07 \AA~ in GA
indicating that GA remains approximately un-rippled even at 900 K.
The variation of $q^*$ against $T$ (Fig. \ref{fig14}(b)) also shows
the same almost constant behavior for GA, whereas for GE, $q^*(T)$
behaves as expected for a 2D membrane~\cite{bilayer}.
From the calculated $H(q\to \infty)$ at different $T$ and Eq. (\ref{hq1}) one can
also determine the $T$ dependence of the bending rigidity $\kappa $.
Using the REBO potential it was found that $\kappa$ decreases with $T$ for GE~\cite{liu},
similarly as for liquid membranes. However, the opposite behavior was
found using the LCBOPII potential~\cite{bilayer}.
In Ref.~\onlinecite{wang} the rigidity was found to depend on the system size.
Thus, the reported values for $\kappa$ in
GE vary from $0.79$ to $2.13$ eV depending on the calculation method~\cite{kars}.
%
In Fig.~\ref{fig14}(c) we show $\kappa$ for GE and GA
calculated from the harmonic part of $H(q)$ between $q$=0.5 $\AA^{-1}$
and $q$=1 $\AA^{-1}$, confirming that for GE, $ \kappa $ increased with $T$.
For GA, $\kappa$ is much larger and
most surprisingly, $ \kappa $ strongly decays when
temperature is increased, opposite to the behavior for GE.

\begin{figure}[t]
\includegraphics[width=0.41\textwidth]{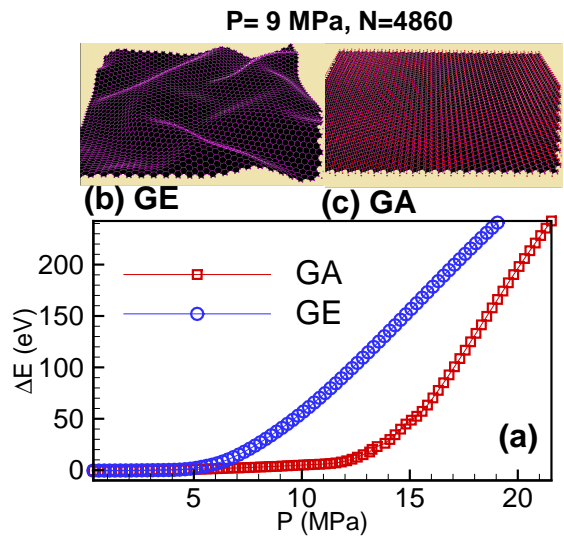}
\caption{(a) Energy increase under bi-axial pressure
with respect to the relaxed system at zero pressure for GA and GE ($0K$).
Shape of (b) GE and (c) GA for P=9 MPa.
Note that while GE is corrugated, GA instead remains un-rippled.
}
\label{fig15}
\end{figure}

The reason why GA does not obey membrane theory
should be found in the geometry of the buckled carbon layer which
allows for low energy in-plane bending modes, involving the C-C-C angles.
These accordion-type of modes have a relatively low energy.
In principle this comes down to a strong anharmonic coupling
of the out-of-plane bending mode with these in-plane bending accordion modes
which strongly damp the out-of-plane excitations. To illustrate
this further we have performed NPT simulations for increasing
pressure, of which the results are shown in Fig.~\ref{fig15}.  It shows
that GA resists much higher pressures before bending than GE~\cite{18}.

Using MD simulations we have shown that the intrinsic thermal ripples
present in GE do not appear in GA for temperatures up to at least 900 $K$,
which we ascribe to the angstrom scale buckling of the carbon layer in GA into
a carbon bilayer-like configuration.
The rippling behavior of GA is in disagreement with
the continuum elasticity theory of membranes.
The results from membrane theory are supposed to be universal
which means that they should not depend on the atomic scale details
within the membrane.
Instead, we find that GA can accommodate the thermal energy by in-plane bending modes, i.e.
modes involving C-C-C bond angles in the buckled carbon layer instead
of leading to significant out-of-plane fluctuations that occur in graphene.
The present results for GA suggests that
membranes of atomic scale thickness can exhibit a more complicated
behaviour than predicted by membrane theory.


{\textit{Acknowledgments}}. \label{agradecimientos}
We thank A. Fasolino, A. Dobry and K. H. Michel for their useful comments.
S. Costamagna is supported by the Belgian Science Foundation (BELSPO). This
work is supported by the ESF-EuroGRAPHENE project CONGRAN and the Flemish
Science Foundation (FWO-Vl).




\begin{thebibliography}{99}

\bibitem{sluiter-sofo}
M. H. F. Sluiter and Y. Kawazoe, Phys. Rev. B {\bf 68}, 085410
(2003). %
J. O. Sofo, A. S. Chaudhari, and G. D. Barber, Phys. Rev. B {\bf
75}, 153401 (2007).

\bibitem{elias}
D. C. Elias, R. R. Nair,T. M. G. Mohiuddin, S. V. Morozov, P. Blake,
M. P. Halsall, A. C. Ferrari, D. W. Boukhvalov, M. I. Katsnelson, A.
K. Geim, and K. S. Novoselov, Science {\bf 323}, 610 (2009).

\bibitem{int2}
M. Z. S. Flores, P. A. S. Autreto, S. B. Legoas and D. S. Galvao,
Nanotechnology {\bf 20}, 465704 (2009). %
%
O. Leenaerts, H. Peelaers, A. D. Hern\'andez-Nieves, B. Partoens,
and F. M. Peeters, Phys. Rev. B {\bf 82}, 195436 (2010).
%
A. D. Hern\'andez-Nieves, B. Partoens, and F. M. Peeters, Phys. Rev.
B {\bf 82}, 165412 (2010). H. Sahin, C. Ataca, and S. Ciraci, Phys. Rev. B  {\bf 81}, 205417 (2010).
Wen, X.D., L. Hand, V. Labet, T. Yang, R. Hoffmann, N.W. Ashcroft,
A. Oganov, and A. Lyakhov, Proc. Natl. Acad. Sci. {\bf 108},
6833 (2011).


\bibitem{mehdi}
M. Neek-Amal and F. M. Peeters, Phys. Rev. B {\bf 83}, 235437
(2011).

\bibitem{PRB77}D. W. Boukhavalov, M. I. Katsnelson, and A. I. Lichtenstein,
Phys. Rev. B {\bf 77}, 035427 (2008).

\bibitem{book2d}D. Nelson, T. Piran and S. Weinberg,
{\it Statistical Mechanics of Membranes and Surface } (Word Scientific,
Singapore, 2004).

\bibitem{fas1}
J. H. Los, M. I. Katsnelson, O. V. Yazyev, K. V. Zakharchenko, and A. Fasolino,
Phys. Rev. B {\bf 80}, 121405 (2009).

\bibitem{GE2d} A. Fasolino, J. H. Los, and M. I. Katsnelson, Nat. Mater. {\bf 6}, 858 (2007).
S. Costamagna and A. Dobry, Phys. Rev. B {\bf 83}, 233401 (2011).
R. Rold\'an, A. Fasolino, K. V. Zakharchenko, and M. I. Katsnelson,
Phys. Rev. B {\bf 83}, 174104 (2011).
M. Neek-Amal, and F. M. Peeters,
Phys. Rev. B {\bf 82}, 085432 (2010).
M. Neek-Amal, and F. M. Peeters,
Appl. Phys. Lett. {\bf 97}, 153118 (2010).






\bibitem{bilayer}
K. V. Zakharchenko, J. H. Los, M. I. Katsnelson, and A. Fasolino,
Phys. Rev. B {\bf 81}, 235439 (2010).

\bibitem{3}  J. C. Meyer, A. K. Geim, M. I. Katsnelson, K. S. Novoselov,
T. J. Booth, and S. Roth, Nature (London) {\bf 446}, 60 (2007).
D. A. Kirilenko, A. T. Dideykin, and G. Van Tendeloo,
Phys. Rev. B {\bf 84}, 235417 (2011).

\bibitem{ledousal}P. Le Doussal and L. Radzihovsky,
Phys. Rev. Lett. {\bf 69}, 1209 (1992).

\bibitem{LCBOPII} J. H. Los, L. M. Ghiringhelli, E. J. Meijer, and A. Fasolino,
Phys. Rev. B {\bf 72}, 214102 (2005).


\bibitem{lammps}$http://lammps.sandia.gov$

\bibitem{AIREBO}S. J.  Stuart, A. B. Tutein, and J. A. Harrison, J. Chem. Phys. {\bf112},
6472 (2000).


\bibitem{liu}P. Liu and Y. W. Zhang, Appl. Phys. Lett. {\bf 94}, 231912 (2009).



\bibitem{wang}
Q. Wang, Phys. Lett. A {\bf 374}, 1180 (2010).


\bibitem{kars}
L. J. Karssemeijer and A. Fasolino, Surf. Sci. {\bf 605}, 1611
(2011). K. H. Michel and B. Verberck,
Phys. Rev. B {\bf 78}, 085424 (2008).
A. Lajevardipour, M. Neek-Amal, and F. M. Peeters,
J. Phys.: Condens. Matter {\bf 24}, 175303 (2012).


\bibitem{18}
See Supplemental Material at [URL will be inserted]




\end{thebibliography}
\end{document}